\documentclass[notoc]{JHEP3}
\usepackage{axodraw}
\usepackage{graphicx}

\def\eps{\epsilon}
\def\a{\alpha}
\def\b{\beta}

\def\gluino#1{
\begin{picture}(160,80)(0,#1)
  \Gluon(15,20)(40,20){2.5}{3}
  \Gluon(15,60)(40,60){2.5}{3}
  \Gluon(40,60)(85,20){-2.5}{10}
  \ArrowLine(85,20)(40,60)
  \Gluon(40,60)(85,60){2.5}{8}
  \ArrowLine(40,60)(85,60)
  \DashLine(40,20)(85,60){4}
  \DashLine(85,20)(40,20){4}
  \ArrowLine(85,60)(130,40)
  \ArrowLine(130,40)(85,20)
  \DashLine(130,40)(150,40){3}
  \Text(62,18)[t]{$\tilde{q}$}
  \Text(62,68)[b]{$\tilde{g}$}
  \Text(110,28)[t]{$q$}
\end{picture}
}

\def\diagtwo#1{
\begin{picture}(160,80)(0,#1)
  \Gluon(15,20)(40,20){2.5}{3}
  \Gluon(15,60)(40,60){2.5}{3}
  \Gluon(40,60)(85,20){-2.5}{10}
  \Gluon(40,60)(85,60){2.5}{8}
  \ArrowLine(40,20)(85,60)
  \ArrowLine(85,20)(40,20)
  \ArrowLine(85,60)(130,40)
  \ArrowLine(130,40)(85,20)
  \DashLine(130,40)(150,40){3}
\end{picture}
}

\title{Evaluating multi-loop Feynman diagrams with infrared and threshold singularities numerically}

\preprint{CERN-PH-TH/2007-058}

\author{Charalampos Anastasiou\\TH Unit, PH Department CERN,
CH-1211 Geneva 23\,, Switzerland\\E-mail: \email{babis@cern.ch}}
\author{Stefan Beerli and Alejandro Daleo\\
Institute for Theoretical Physics,ETH,
CH-8093, Zurich\,, Switzerland\\E-mail: \email{sbeerli@itp.phys.ethz.ch}\\E-mail: \email{adaleo@itp.phys.ethz.ch}\\}

\abstract{We present a method to evaluate numerically Feynman diagrams 
directly from their Feynman parameters representation. We first disentangle 
overlapping singularities using sector decomposition. Threshold 
singularities are treated with an appropriate contour 
deformation. We have validated our technique comparing with recent 
analytic results for the $gg \to h$ two-loop amplitudes with heavy quarks 
and scalar quarks.}

\keywords{QCD, NLO and NNLO Computations}

\begin{document}
\section{Introduction}
\label{sec:introduction}
The evaluation of complicated Feynman diagrams remains one of the theoretical 
challenges to be met for the needs of the ongoing collider physics program. 
We develop an automated method to evaluate  loop diagrams with infrared and 
threshold  singularities in all kinematic regions. The method combines  
sector decomposition to simplify their infrared singular structures 
and a deformation of the integration path to treat threshold singularities. 

Sector decomposition was introduced as a  simple systematic algorithm to 
evaluate loop integrals by Binoth and 
Heinrich~\cite{Binoth:2000ps,Binoth:2003ak}.   
The algorithm divides iteratively the integration region into sectors;  in each
sector the integration variables which could produce an overlapping 
singularity are ordered according to their magnitude. 
The overlapping singularity takes the form of a pole in  
the variable which approaches the singular limit first and 
can be factored out.    

A concern for the viability of the 
method has been the proliferation of terms.  In 
explicit calculations of cross-sections through 
next-to-next-to-leading order, it has 
been shown that one can write efficient sector decomposition algorithms  
for realistic applications in gauge field 
theories~\cite{Melnikov:2006kv,Melnikov:2006di,Anastasiou:2005pn,Anastasiou:2005qj,Anastasiou:2004xq,Anastasiou:2004qd}. 
However, to date, automated sector decomposition is limited to the 
calculation of infrared divergent loop diagrams in kinematic regions with 
trivial or no thresholds. This could halt progress 
in evaluating amplitudes for interesting scattering processes 
and  fusion processes via heavy particles.   

During the last few years an inspired method is being developed for the 
evaluation of one-loop amplitudes by Nagy and 
Soper~\cite{Nagy:2003qn,Nagy:2006xy}. In their method, the  infrared and 
ultraviolet divergences are matched algorithmically by  simple counterterms 
for each diagram; these add up to  
functions which integrate to the known universal poles in $\epsilon$ 
of one-loop amplitudes.  After the one-loop integrals are rendered finite 
in four dimensions, they perform a numerical integration over Feynman 
parameters and the loop momentum.   
They have proposed a systematic way to find an integration contour in the 
space of Feynman parameters which is suitable for a numerical integration.

We have merged the algorithm for sector decomposition with the contour 
deformation of Feynman parameters proposed by Nagy and 
Soper~\cite{Nagy:2006xy}. 
Since sector decomposition offers a general solution to rendering Feynman 
diagrams with divergences in $d \to 4$ dimensions finite, in principle, we can now compute generic 
multi-loop integrals numerically in all kinematic regions. 

We have written three independent computer implementations of our 
method and performed extensive checks evaluating a variety of one and 
two-loop scalar and tensor integrals which we could verify with other 
methods. To prove the efficiency of our method, we have recomputed all 
diagrams in the two-loop amplitudes for $gg \to h$ production via heavy 
quarks and squarks,  recently computed analytically.  The new method yields 
numerical results which  are in excellent agreement with the analytic 
evaluation~\cite{Anastasiou:2006hc}. 

Lazopoulos, Melnikov and  Petriello have recently 
presented~\cite{Lazopoulos:2007} the evaluation of the NLO QCD 
corrections to $pp \to ZZZ$. The method in their publication is the 
same as the one we are presenting. Here, we are applying it to the evaluation 
of a two-loop amplitude; together with the work of~\cite{Lazopoulos:2007}
this emphasizes further the versatility of the method.     

Binoth et al.  have used a contour 
deformation to evaluate integrals which are free of infrared 
divergences~\cite{Binoth:2005ff}. They also noted the possibility 
of merging sector decomposition  and contour deformation as an alternative 
to their reduction method. To our understanding, the viability of 
this idea  for practical applications was not  investigated 
in~\cite{Binoth:2005ff}.

A competitive numerical method for the evaluation of loop 
amplitudes is via Mellin-Barnes 
representations~\cite{Anastasiou:2005cb,Czakon:2005rk}. Loop 
integrals with a small number of kinematic scales  tend to 
have Mellin-Barnes representations with a lower dimensionality 
than the corresponding Feynman parameter representations. 
In such cases the Mellin-Barnes method should be advantageous; 
however, the method of this paper could perform better by increasing 
the number  of kinematic scales. In addition, 
as noted in~\cite{Czakon:2005rk}, Mellin-Barnes integrals cannot be computed 
stochastically  in phase-space regions with mass dependent thresholds.  
For such  applications the Mellin-Barnes method can be used 
only for checking purposes in the Euclidean region; 
sector decomposition with contour deformation could be then the 
only viable numerical method to obtain a physical result.     

We should also note that a very 
significant progress in developing numerically integrable 
representations  for  two-loop three-point functions with threshold and 
infrared  singularities has been made 
in~\cite{Passarino:2006gv}. However, this approach is yet not fully automated 
or general and requires an  indivindual study for each 
topology. 
  
We now present the method and our numerical results and comparisons. 

\section{Method}
\label{sec:method}

We first introduce independent Feynman parameters for a 
multi-loop Feynman diagram. 
In general, this yields a sum of terms of the form
\begin{equation}
\label{eq:startfp}
I = C(\eps) \lim_{\delta \rightarrow 0} 
\int_0^1 dx_1 \cdots dx_n \frac{{\cal F}(\vec{x}, \eps)} 
{\left[ {\cal G}(\vec{x},M_i^2,s_{kl})-i\delta\right]^{\alpha+ n_L \eps}}
\end{equation}
where $\alpha$ is an integer, $n_L$ is the number of loops and $M_i, s_{kl}$ 
are masses and kinematic invariants. The function ${\cal G}$ is a polynomial 
of the independent Feynman parameters $\vec{x}$. 

The integrand can be singular at the edges of the integration region. 
As a first step, we disentangle overlapping 
singularities using sector  
decomposition~\cite{Hepp:1966eg,Roth:1996pd,Binoth:2000ps}. 
The outcome is a sum of integrals of the type
\begin{equation}
\label{eq:sectorfp}
I_{s} = C(\eps) \lim_{\delta \rightarrow 0} 
\int_0^1  \frac{dx_1 \cdots dx_n x_1^{-\a_1+\b_1\eps}\cdots 
x_n^{-\a_n+\b_n\eps} 
{\cal F}_{s}(\vec{x}, \eps)} 
{\left[ {\cal G}_{s}(\vec{x},M_i^2,s_{kl})-i\delta\right]^{\alpha+ n_L \eps}}
\end{equation}
The singularities at the edges of the  integration region 
are now factorized. The function ${\cal G}_{s}$ is finite 
at $x_i \to 0,1$ and the singularities from the factors $x^{-\a+\b\eps}$ when 
$\a>0$ can be extracted independently for each integration variable. 
However, ${\cal G}_{s}$ may produce singularities 
if it vanishes inside the integration region. It is important 
to postpone the extraction of the $\epsilon$ poles until we treat 
these threshold singularities first.

Following the method of Nagy and Soper~\cite{Nagy:2006xy}, 
we construct a contour of integration where the imaginary part of 
${\cal G}_{s}$ is negative, that is, enforcing the $-i\,\delta$ 
prescription already present in the original integral. 
The new contour is defined by deforming the integration path of every 
Feynman parameter; it is parameterized by
\begin{equation}
\label{eq:deformation}
z_i = x_i - i\lambda x_i (1-x_i)\frac{\partial {\cal G}_s}{\partial x_i}\,. 
\end{equation}
Provided that no poles were crossed in going from $[0,1]$ to the contour $C$, we have
\begin{equation} 
\int_0^1 \frac{\left(\prod_{j=1}^n dx_j x_j^{-\a_j+\b_j\eps}\right)\,{\cal F}_{s}(\vec{x}, \eps)} 
{\left[ {\cal G}_{s}(\vec{x},M_i^2,s_{kl})-i\delta\right]^{\alpha+ n_L \eps}}
= 
\int_C  \frac{\left(\prod_{j=1}^n dz_j z_j^{-\a_j+\b_j\eps}\right)\,
{\cal F}_{s}(\vec{z}, \eps)} 
{\left[ {\cal G}_{s}(\vec{z},M_i^2,s_{kl})\right]^{\alpha+ n_L \eps}}\,.
\end{equation} 
The choice of Eq.~(\ref{eq:deformation}) for the contour deformation guarantees that for small values of $\lambda$, 
the function ${\cal G}_{s}$ acquires a negative 
imaginary part of order ${\cal O}(\lambda)$ 
\begin{equation}
\label{eq:imaginaryG}
{\cal G}_{s}(\vec{z}) = {\cal G}_{s}(\vec{x})-i\lambda  
\sum_i x_i (1-x_i) \left( \frac{\partial {\cal G}_{s}}{\partial x_i} \right)^2 
+ {\cal O}(\lambda^2),  
\end{equation}
where the ${\cal O}(\lambda^2)$ terms are purely real.
One could add higher order $\lambda^n$ terms in Eq.~(\ref{eq:deformation}) 
to cancel the imaginary parts of ${\cal G}_s$ at ${\cal O}(\lambda^3)$ and 
higher orders. In practice,
it is sufficient to perform a linear deformation 
as in Eq.~(\ref{eq:deformation}), 
and choose  a small enough value for $\lambda$ such that these 
contributions are suppressed.

Changing variables using the parameterization in 
Eq.~(\ref{eq:deformation}),
each sector integral is now written as 
\begin{eqnarray}
\label{eq:afterdeform}
I_{s}&=&C(\eps) \int_0^1 \prod_{j=1}^n dx_j z_j^{-\a_j+\b_j\eps}\, 
{\cal J}(\vec{x} \to \vec{z}) 
{\cal L}(\vec{z}(\vec{x}),\eps) \nonumber \\
&=& 
C(\eps) \int_0^1 \prod_{j=1}^n dx_j x_j^{-\a_j+\b_j\eps}\,\left(\frac{z_j}{x_j}\right)^{-\a_j+\b_j\eps}\, 
{\cal J}(\vec{x} \to \vec{z}) {\cal L}(\vec{z}(\vec{x}), \eps) 
\end{eqnarray}
where ${\cal J}(\vec{x} \to \vec{z})$ is the Jacobian of the 
transformation of Eq.~(\ref{eq:deformation}). The function ${\cal L}$ 
is finite in the boundary of the integration region and thus 
can be expanded as a Taylor series around $\eps=0$. 
From  Eq.~(\ref{eq:deformation}) we can 
see that the ratios $z_i/x_i = 1 + {\cal O}(x_i)$ are also smooth in 
the singular limits. 

Now, we are free to extract the $\epsilon$ 
poles in each integration variable by applying
\begin{eqnarray}
\label{eq:subtraction}
\int_0^1 dx x^{-n+\eps} f(x)  &=& \int_0^1 dx x^\eps 
\frac{f(x)- \sum_{k=0}^{n-1} x^{k} \frac{f^{(k)}(0)}{k!} }{x^{n}}
+\sum_{k=0}^{n-1}  \frac{f^{(k)}(0)}{k! (k+1-n+\eps)}
\end{eqnarray}
on Eq.~(\ref{eq:afterdeform}). After this expansion, we are left 
with integrals which can be safely expanded in power series in 
$\epsilon$. We compute the coefficients  of the $\epsilon$ 
series numerically. Notice that in presence of 
higher order singularities, $n>1$, the 
expansion in Eq.~(\ref{eq:subtraction}) involves derivatives of both 
the Jacobian and the function ${\cal L}$ that contains all non-singular 
factors, including factors coming from the tensor structure of the
integral. 

There are many options for the numerical evaluation of the resulting 
integrals. For example, we can combine the 
contributions from all sectors into a single integrand; alternatively,
we can integrate each sector separately and sum up the results.  
We have found that the second choice is usually better, since 
the adaptation of numerical integration algorithms is more effective when  
dealing with simpler integrands. 

We have implemented the method described above into three different 
programs. Sector decomposition and contour deformation are performed with 
MAPLE and MATHEMATICA routines. The same programs control the 
creation of numerical routines for the evaluation of the integrals; 
these are written in FORTRAN or C++. In all implementations we use 
the integration routines in the Cuba~\cite{Hahn:2004fe} library, 
relying mostly 
on the Cuhre and Divonne algorithms. 

An important issue when performing the numerical integration is 
the value of the parameter $\lambda$ in Eq.~(\ref{eq:deformation}),
which controls the magnitude of the contour deformation.  
A very small value could result to instabilities due to rounding errors. 
A very big value could result to a deformation with the wrong 
sign in Eq.~(\ref{eq:deformation}). This last case is easy to detect at
runtime and we have implemented diagnostic routines to abort the 
numerical evaluation if ${\cal G}_{s}$ is found to have an imaginary part 
with the wrong sign. As we will show in the next section in
an specific example, there is usually a very comfortable interval 
for $\lambda$  where the result of the integration is insensitive to 
its value.

\section{Results}
\label{sec:results}

We  have applied our method to compute the two-loop amplitudes 
for $gg\to h$ mediated by a heavy quark or a scalar quark purely 
numerically. These amplitudes have been computed earlier 
either by using a mixture of analytic and numerical 
integrations~\cite{Spira:1995rr,Muhlleitner:2006wx}
or analytically in ~\cite{Anastasiou:2006hc} and 
in~\cite{Aglietti:2006tp}~\footnote{ 
The integral representation of the result in~\cite{Spira:1995rr} was 
expressed in an analytic form in~\cite{Harlander:2005rq}}.
\FIGURE[h]{
\begin{tabular}{cc}
\diagtwo{0}&\gluino{0}\\
(a)&(b)
\end{tabular}
\caption{(a) Feynman diagram contributing to $gg\to h$ with a heavy quark
  loop. (b) Master integral arising in the calculation of $gg\to h$ in the
  MSSM.}\label{fig:diagrams}}
We have evaluated all Feynman diagrams in these amplitudes with a very 
good numerical precision. As an example, in Fig.~\ref{fig:diag2} 
\FIGURE[h]{
\includegraphics[height=7cm]{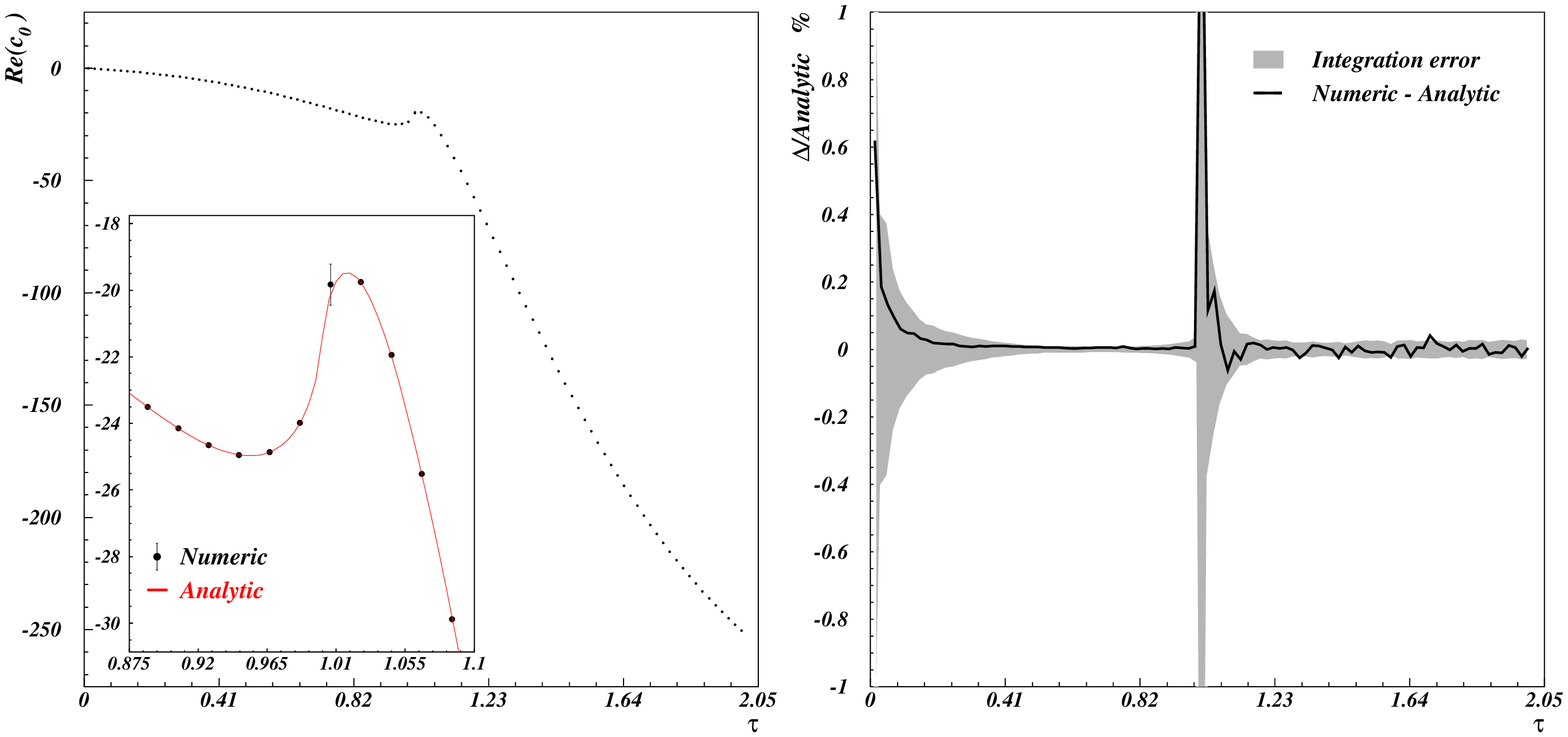}
\caption{Results for the real part of the finite piece ($\mbox{Re}(c_0)$) 
of the 
Feynman diagram in Fig.~\ref{fig:diagrams}a. The left
panel shows the results of the numerical integration as black dots with error
bars. The inset plot zooms in on the threshold region, the red line 
corresponds to the evaluation of the analytic result of
~\cite{Anastasiou:2006hc}. The right panel shows the difference in percent 
of the numerical evaluation and the
analytic one, normalized to the latter. The gray bands correspond to the
integration error. At threshold this error is 3\%.}\label{fig:diag2}}
we show our results for the diagram in Fig.~\ref{fig:diagrams}a. On
the left panel we plot the real part of the finite piece of the diagram as
a function of $\tau=s/(4\,m^2)$, normalized to $m^2=1$. The inset plot
gives a more detailed view of the threshold region, where the numerical
integration is most difficult, superimposed with the analytic results 
from~\cite{Anastasiou:2006hc}. The plot on the
right panel shows the percent difference between the numerical result and
the analytic one, normalized to the analytic value. In black lines we included
the bands corresponding to the integration error, obtained by adding in
quadrature the errors quoted by the integration routine for each
sector. We obtain similar results for the single pole and for the imaginary
parts of both the single pole and the finite piece of this diagram.

As stressed above, the method relies on the proper choice of the value for the
parameter $\lambda$. Values that are too large produce imaginary parts with
the wrong sign for the function ${\cal G}_{s}$. Very small values generate a
contour that is too close to the real line --and thus to the zeros of ${\cal
  G}_{s}$-- and produce numerical instabilities. 
In practical implementations, we found that
there is usually a good range for $\lambda$. As an example, in
Fig.~\ref{fig:lambda} we plot the results for the scalar integral
corresponding to the diagram in Fig.~\ref{fig:diagrams}a as a function of
$\lambda$ for two different values of $\tau$.
\FIGURE[h]{
\includegraphics[height=6cm]{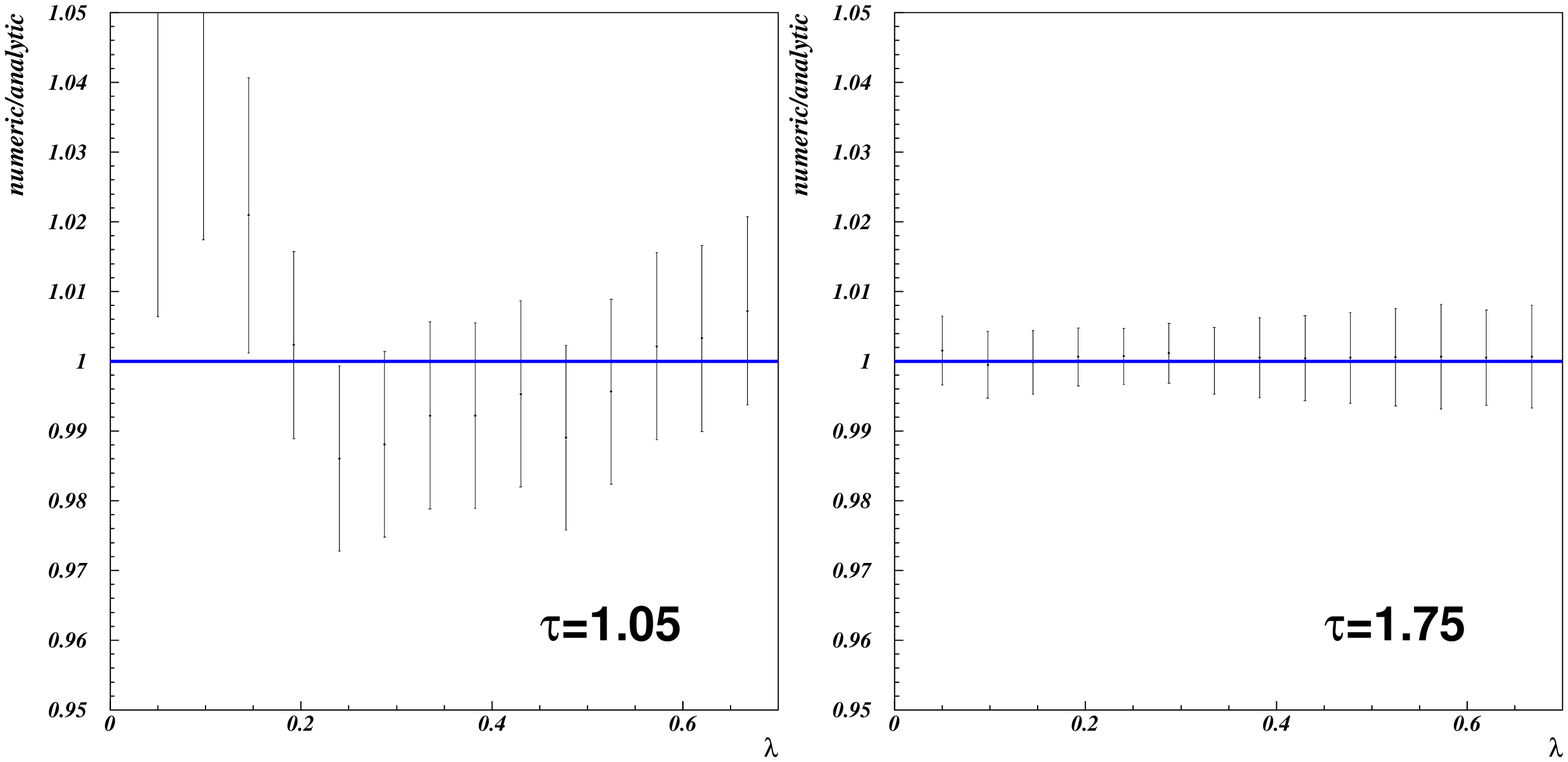}
\caption{Results for the real part of the finite piece of the scalar integral 
corresponding to the Feynman diagram in Fig.~\ref{fig:diagrams}a as a function
of the parameter $\lambda$ for two fixed values of the kinematical ratio
$\tau$. The results have been normalized to the analytic
result for this master integral obtained in ~\cite{Anastasiou:2006hc}. 
}\label{fig:lambda}
}
The results in these plots show that away from the threshold region, the
integration is rather insensitive to the value of $\lambda$ chosen, as long as
it induces a deformation providing the right sign for the imaginary part in 
Eq.~(\ref{eq:imaginaryG}). On the other hand, close to the threshold region,
the magnitude of the deformation has to be larger in order to get a reliable
estimate of the integral.  

As a novel result, we applied our method to the calculation of the scalar
integral corresponding to the Feynman diagram of Fig.~\ref{fig:diagrams}b. 
This integral is one of the
master integrals appearing in the SUSY QCD corrections to $gg\to h$ and it
involves a massive quark, a massive scalar quark and a massive gluino. 
\FIGURE[h]{
\includegraphics[height=6cm]{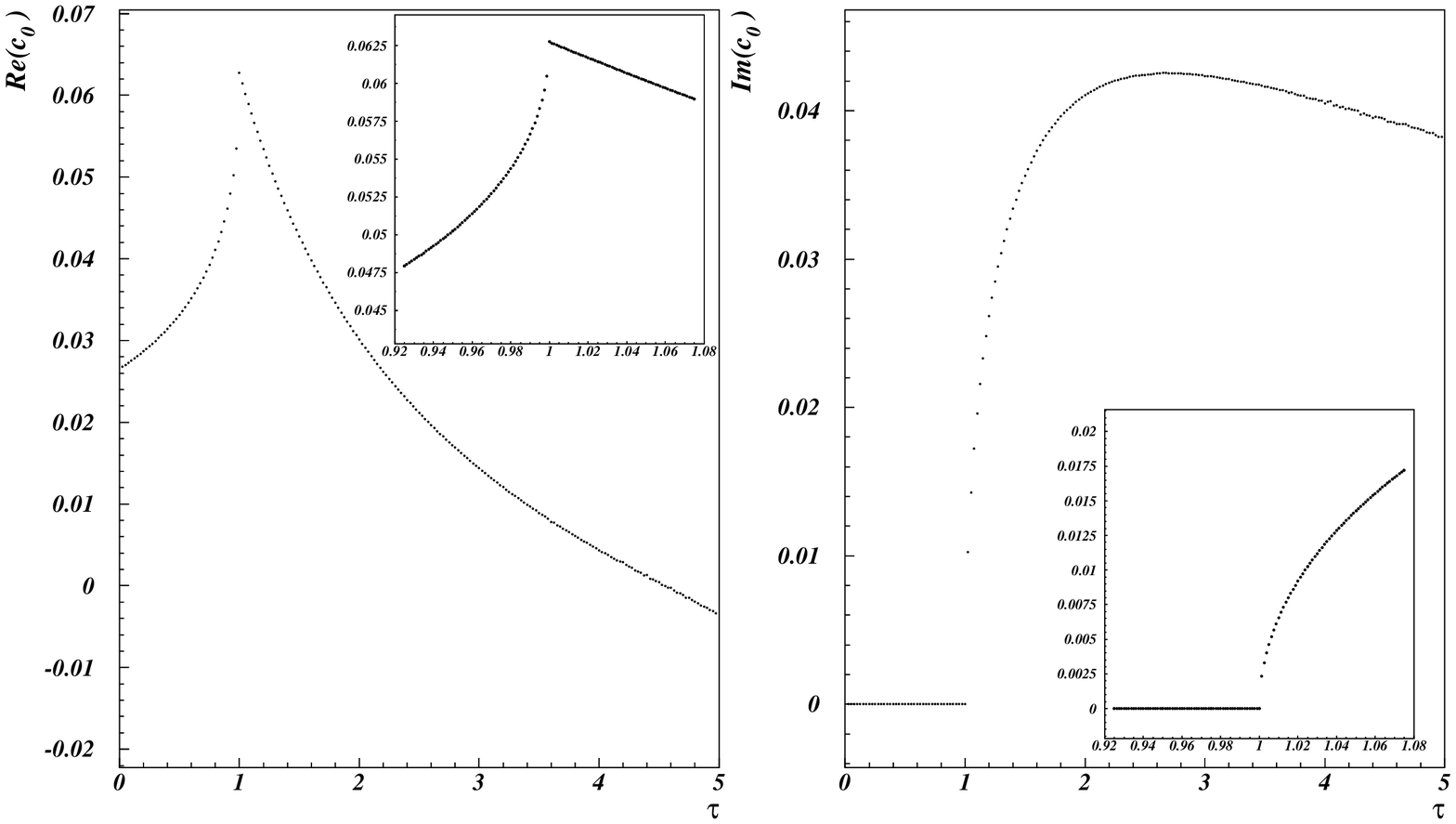}
\caption{Results for the scalar integral 
corresponding to the Feynman diagram in Fig.~\ref{fig:diagrams}b as a function
of $\tau=s/(4m_q^2)$ for fixed values of  $m_{\tilde{g}}^2=400/175\,m_q^2$ and
$m_{\tilde{q}}^2=600/175\,m_q^2$. The inset plots zoom in the threshold
region. The estimated relative accuracy of the points is better that 1 per mille.}\label{fig:gl1}
}
Our results are displayed in Fig.~\ref{fig:gl1} as a function of
$\tau=s/(4m_q^2)$ for fixed values of $m_{\tilde{g}}^2=400/175\,m_q^2$ and
$m_{\tilde{q}}^2=600/175\,m_q^2$ with $m_q=1$. The results are again very
stable over a wide range of $\lambda$. Due to the absence of massless
propagators, the numerical evaluation of this integral turns out to be
substantially faster than the scalar integral corresponding to 
the diagram of Fig.~\ref{fig:diagrams}a.

\section{Conclusions}
\label{sec:conclusions}  
We present a new method for the numerical evaluation of 
multi-loop Feynman diagrams containing both infrared and threshold
singularities. The method uses sector decomposition to 
extract the infrared singularities followed by contour deformation
in the Feynman parameters to deal with the thresholds present in 
the diagram. The algorithmic nature of the approach naturally leads to
a high degree of automatization in all the stages of the calculation. 

We tested the method recalculating the two loop 
corrections to $gg\to h$ mediated by a massive quark and a massive scalar 
quark. We find that the method is very efficient and reliable, reproducing
the analytic results with great accuracy. 

Currently we are applying the technique to the calculation of the two
loop SUSY QCD corrections to $gg\to h$. As an example, we have presented 
here results for one of the most complicated with analytical methods, 
yet uncalculated, master integrals appearing in this amplitude. 
We find excellent numerical behavior, showing that the framework has 
a great potential for computing automatically
general multi-loop processes involving internal 
thresholds.

\acknowledgments
We are grateful to Uli Haisch and Zoltan Kunszt 
for many useful discussions and their encouragement.  We thank Achilleas 
Lazopoulos, Kirill Melnikov and Frank Petriello for informing us of their 
publication~\cite{Lazopoulos:2007} before submiting to the archive. The work
of Stefan Beerli and Alejandro Daleo was supported by the Swiss
National Science Foundation (SNF) under contract number
200020-113567/1.



\begin{thebibliography}{999}

\bibitem{Binoth:2000ps}
  T.~Binoth and G.~Heinrich,
  Nucl.\ Phys.\  B {\bf 585}, 741 (2000)
  [arXiv:hep-ph/0004013].

\bibitem{Binoth:2003ak}
  T.~Binoth and G.~Heinrich,
  Nucl.\ Phys.\  B {\bf 680}, 375 (2004)
  [arXiv:hep-ph/0305234].


\bibitem{Melnikov:2006kv}
  K.~Melnikov and F.~Petriello,
  Phys.\ Rev.\  D {\bf 74}, 114017 (2006)
  [arXiv:hep-ph/0609070].

\bibitem{Melnikov:2006di}
  K.~Melnikov and F.~Petriello,
  Phys.\ Rev.\ Lett.\  {\bf 96}, 231803 (2006)
  [arXiv:hep-ph/0603182].


\bibitem{Anastasiou:2005pn}
  C.~Anastasiou, K.~Melnikov and F.~Petriello,
  arXiv:hep-ph/0505069.

\bibitem{Anastasiou:2005qj}
  C.~Anastasiou, K.~Melnikov and F.~Petriello,
  Nucl.\ Phys.\  B {\bf 724}, 197 (2005)
  [arXiv:hep-ph/0501130].

\bibitem{Anastasiou:2004xq}
  C.~Anastasiou, K.~Melnikov and F.~Petriello,
  Phys.\ Rev.\ Lett.\  {\bf 93}, 262002 (2004)
  [arXiv:hep-ph/0409088].

\bibitem{Anastasiou:2004qd}
  C.~Anastasiou, K.~Melnikov and F.~Petriello,
  Phys.\ Rev.\ Lett.\  {\bf 93}, 032002 (2004)
  [arXiv:hep-ph/0402280].

\bibitem{Nagy:2003qn}
  Z.~Nagy and D.~E.~Soper,
  JHEP {\bf 0309}, 055 (2003)
  [arXiv:hep-ph/0308127].

\bibitem{Nagy:2006xy}
  Z.~Nagy and D.~E.~Soper,
  Phys.\ Rev.\  D {\bf 74}, 093006 (2006)
  [arXiv:hep-ph/0610028].

\bibitem{Anastasiou:2006hc}
  C.~Anastasiou, S.~Beerli, S.~Bucherer, A.~Daleo and Z.~Kunszt,
  JHEP {\bf 0701}, 082 (2007)
  [arXiv:hep-ph/0611236].


\bibitem{Lazopoulos:2007}
  A.~Lazopoulos, K.~Melnikov and F.~Petriello,
  arXiv:hep-ph/0703273.

\bibitem{Binoth:2005ff}
  T.~Binoth, J.~P.~Guillet, G.~Heinrich, E.~Pilon and C.~Schubert,
  JHEP {\bf 0510}, 015 (2005)
  [arXiv:hep-ph/0504267].


\bibitem{Anastasiou:2005cb}
C.~Anastasiou and A.~Daleo,
JHEP {\bf 0610}, 031 (2006)
[arXiv:hep-ph/0511176].


\bibitem{Czakon:2005rk}
  M.~Czakon,
  Comput.\ Phys.\ Commun.\  {\bf 175}, 559 (2006)
  [arXiv:hep-ph/0511200].

\bibitem{Passarino:2006gv}
  G.~Passarino and S.~Uccirati,
  Nucl.\ Phys.\  B {\bf 747}, 113 (2006)
  [arXiv:hep-ph/0603121].


\bibitem{Hepp:1966eg}
  K.~Hepp,
  Commun.\ Math.\ Phys.\  {\bf 2}, 301 (1966).

\bibitem{Roth:1996pd}
  M.~Roth and A.~Denner,
  Nucl.\ Phys.\  B {\bf 479}, 495 (1996)
  [arXiv:hep-ph/9605420].

\bibitem{Hahn:2004fe}
  T.~Hahn,
  Comput.\ Phys.\ Commun.\  {\bf 168}, 78 (2005)
  [arXiv:hep-ph/0404043].

\bibitem{Spira:1995rr}
  M.~Spira, A.~Djouadi, D.~Graudenz and P.~M.~Zerwas,
  Nucl.\ Phys.\  B {\bf 453}, 17 (1995)
  [arXiv:hep-ph/9504378].

\bibitem{Muhlleitner:2006wx}
  M.~Muhlleitner and M.~Spira,
  arXiv:hep-ph/0612254.

\bibitem{Aglietti:2006tp}
  U.~Aglietti, R.~Bonciani, G.~Degrassi and A.~Vicini,
  JHEP {\bf 0701}, 021 (2007)
  [arXiv:hep-ph/0611266].

\bibitem{Harlander:2005rq}
  R.~Harlander and P.~Kant,
  JHEP {\bf 0512}, 015 (2005)
  [arXiv:hep-ph/0509189].


\end{thebibliography}
\end{document}